\documentclass[twocolumn,aps,showpacs]{revtex4}
\usepackage{graphicx,color}

\begin{document}
\title{High contrast Mach-Zehnder lithium atom interferometer in the Bragg regime}

\author{ R. Delhuille, C. Champenois \footnote{present address: Physique des Interactions
Ioniques et Mol\'eculaires, Universit\'e de Provence et CNRS, UMR
6633, 13397 Marseille cedex 20, France}, M. B\"uchner, L.
Jozefowski,\\ C. Rizzo, G. Tr\'enec and J. Vigu\'e}
\affiliation{ Laboratoire Collisions Agr\'egats R\'eactivit\'e-IRSAMC
\\Universit\'e Paul Sabatier and CNRS UMR 5589
\\ 118, Route de Narbonne; 31062 Toulouse Cedex, France
\\ Fax : ~{\tt 33 5 61 55 83 17}
\\ e-mail :~{\tt jacques.vigue@irsamc.ups-tlse.fr}}

%\date{\today}

\begin{abstract}
We have constructed an atom interferometer of the Mach-Zehnder
type, operating with a supersonic beam of lithium. Atom
diffraction uses Bragg diffraction on laser standing waves. With
first order diffraction, our apparatus has given a large signal
and a very good fringe contrast ($74$\%), which we believe to be
the highest ever observed with thermal atom interferometers. This
apparatus will be applied to high sensitivity measurements.

%PACS : 03.75 Dg, 32.80.Lg, 39.20.+q
\end{abstract}

\pacs{03.75.Dg,32.80.Lg,39.20.+q}

\maketitle

%%%%%%%%%%%%%%%%%%%%%%%%%%%%%%%%%%%%%%%%%%%%%%%%%%%%%%%%%%%%%%%%%%%%%%

Several different atom interferometers gave their first signals in
1991 :
\begin{itemize}
\item a Young's double slit experiment was demonstrated by O. Carnal
and J. Mlynek, with a supersonic beam of metastable helium
\cite{carnal91}
\item a Mach-Zehnder interferometer built by D. Pritchard and coworkers
using a thermal atomic beam of sodium and diffraction on material
gratings \cite{keith91}
\item  an interferometer based on Ramsey fringes in saturated
absorption spectroscopy, following the idea of Ch. Bord\'e
\cite{borde89}, was built by J. Helmcke and coworkers, with a
thermal atomic beam of calcium and it was used to demonstrate
Sagnac effect with atomic waves \cite{riehle91}
\item an interferometer using laser cooled sodium atom and Raman
diffraction was built by M. Kasevich and S. Chu and gave the first
high sensitivity measurement of the local acceleration of gravity
based on atom interferometry \cite{kasevich91}
\end{itemize}
This research field has developed rapidly since 1991 and an
excellent overview of this field and of its applications can be
found in the book "Atom interferometry" \cite{berman97}.

In this paper, we describe the first interference signals observed
with our newly built Mach-Zehnder atom interferometer operating
with thermal lithium atoms. The diffraction gratings, which are
used as mirrors and beam-splitters, are made of laser standing
waves operating in the Bragg regime. Our first signals present a
very good signal to noise ratio, a mean detected atom flux of $1.4
\times 10^4$ $s^{-1}$ and a $74$\% fringe contrast. As far as we
know, this is the highest contrast ever observed with a thermal
atom Mach-Zehnder interferometer.

Let us recall the performances achieved by this family of atom
interferometers. In each case, we give the mean value of the
detected atomic flux $I$ and the fringe contrast (or visibility)
${\mathcal{C}}$. These parameters are both important for phase
measurements : assuming a Poisson statistics for the noise, the
accuracy of these measurements increases with a figure of merit
given by $I{\mathcal{C}}^2$. In 1991, the interferometer of D.
Pritchard and coworkers \cite{keith91} gave a $13$\% contrast with
a mean detected atom flux of $290$ $s^{-1}$, values improved in
1997 up to a $49$\% contrast and a mean flux of $1900$ $s^{-1}$
\cite{schmiedmayer97} or a $17$\% contrast and a mean flux of
$2\times10^5$ $s^{-1}$\cite{lenef97}. In 1995, A. Zeilinger and
coworkers \cite{rasel95} operated an interferometer using
metastable argon and laser diffraction in the Raman-Nath regime
which produced a $10$\% contrast associated to a mean detected
flux of $1.4\times 10^4$ $s^{-1}$. Also in 1995, S. A. Lee and
coworkers \cite{giltner95b} built an interferometer using
metastable neon and laser diffraction in the Bragg regime and they
observed a $62$\% contrast associated to a mean detected flux of
$1.5\times 10^3$ $s^{-1}$. Finally, in 2001, an helium
interferometer built by J.P. Toennies and coworkers
\cite{toennies01}, with material gratings, has given a $71$ \%
contrast with a mean detected flux close to $10^3$ $s^{-1}$.

We have limited the present comparison to the family of
interferometers which rely on elastic diffraction, i.e. in which
the atom internal state is not modified by the diffraction
process. However, as discussed by Ch. Bord\'e
\cite{borde89,borde97}, the general case is inelastic diffraction,
which is used in Ramsey-Bord\'e interferometers \cite{riehle91}
and also in Mach-Zehnder atom interferometers. This type of
interferometer can provide a very high output flux because the
various outputs are distinguished by the atom internal state and
not only by the direction of propagation : this circumstance
permits to use a broad (but well collimated) atomic beam. One of
the best examples is the cesium interferometer developed by
Kasevich and coworkers as a gyroscope of extremely high
sensitivity. This interferometer uses a thermal atomic beam of
cesium, with transverse laser cooling and it has produced a fringe
contrast of $20$\% \cite{gustavson97}, an output flux equal
$1\times 10^{11}$ $s^{-1}$, and a signal to noise ratio of $33
000$ for $1$ s of integration \cite{gustavson00}. However, this
advantage of Raman interferometers is obtained only if one does
not separate the atomic paths in order to apply a perturbation to
one of the two paths. This limitation is one of the reasons which
explain why we have not chosen to develop a Raman interferometer.

When building an atom interferometer, two very important choices
must be made, namely the atom and the diffraction process. The
choice of the atom is largely limited by the possibility of
producing either a very intense atomic beam or by the availability
of a very high detection sensitivity. For thermal atoms, an
efficient laser induced fluorescence detection is feasible
\cite{gustavson97,gustavson00} but difficult because the time
spent in the laser excitation volume is small. Another detection
technique is based on surface ionization, which is very efficient
only with alkali atoms or with metastable atoms. We have chosen to
work with an alkali atom in its ground state, because the
interactions of these atoms are more accurately described by ab
initio quantum chemistry calculations than those of metastable
rare gas atoms. Then, a very important design parameters is the
first order diffraction angle $\theta_1 = \lambda_{dB}/a = h/(m v
a)$ (where $a$ is the grating period, $m$ the atomic mass and $v$
the velocity) as it defines the needed collimation of the atomic
beam and also the separation of the atomic paths near the second
grating. For supersonic beams of light alkalis seeded in a carrier
gas, the beam velocity $v$ depends almost solely of the carrier
gas molecular mass and a small velocity requires a heavy carrier
gas. A convenient and inexpensive choice is argon, which gives $v$
close to $1050$ m/s, for a temperature $T$ near $1050$ K.

We may compare our choices to those of D. Pritchard, as our
interferometer design is largely inspired by his design. The
choices made by D. Pritchard were to use material gratings with
very small $a$ value, $a = 200$ nm in most experiments (but some
experiments involved smaller values down to $100$ nm) and sodium
atom (molar mass $23$ g), with a de Broglie wavelength
$\lambda_{dB} \approx 17$ pm and a first order diffraction angle
$\theta_1 \approx 85 $ $\mu$rad. As further discussed below, we
have made the choice of using laser diffraction and, in this case,
the grating period is $a =\lambda_r/2$, where $\lambda_r$ is the
wavelength of the resonance transition. Usually, the first
resonance transition is chosen for intensity reasons and practical
considerations (laser cost, power and availability) and the
achieved grating periods are not very small, in the $300-500$ nm
range. For lithium, the $a$ value ($a= 335$ nm corresponding to
$\lambda_r = 671$ nm), substantially larger than the $a$ value
commonly used by D. Pritchard, is compensated by the smaller mass
(molar mass $7$ g). With a de Broglie wavelength
$\lambda_{dB}\approx 54$ pm, the first order diffraction angle is
$\theta_1 \approx 160$ $\mu$rad (from now on, we will discuss only
the case of $^7$Li which represents $92.6$\% of natural lithium
and which is selected by our choice of laser frequency). We want
to take advantage of this relatively large diffraction angle to
make interferometric experiments, with only one atomic path
submitted to a perturbation. Such experiments have been done only
by the group of D. Pritchard \cite{ekstrom95,schmiedmayer95},
using a separation between the two atomic paths of the order of
$55$ $\mu$m near the second grating, while in our apparatus the
corresponding separation is equal to $100$ $\mu$m. Obviously,
considerably larger separation values can be achieved by using a
slow atomic beam, as produced by laser cooling, but, up to now, in
cold atom interferometers, the various atomic paths have not been
separated because of the size of the cold samples.

Laser diffraction of atoms results from the proposal of electron
diffraction by light due to Kapitza and Dirac \cite{kapitza33},
generalized to atom diffraction by S. Altshuler et al.
\cite{altshuler66}. In the Bragg geometry, the oscillating
character of the electron diffraction amplitude appears in the
work of M. Federov \cite{federov67} and R. Gush et al.
\cite{gush71}. Early general theoretical analysis of laser
diffraction of atoms are due to A. Bernhardt and B. Shore
\cite{bernhardt81} as well as to Ch. Tanguy et al.
\cite{tanguy84}. The Rabi-oscillation regime in the Bragg geometry
was discussed by D. Pritchard and P. Gould in 1985
\cite{pritchard85} and observed by the same research group in 1987
\cite{martin88}. The interest of Bragg diffraction for
interferometry is that the incident beam is split almost perfectly
in only two beams \cite{giltner95a,borde97} and the relative
intensities of these two beams can be tuned at will by varying the
laser power density and/or the interaction time. In a perturbation
viewpoint, diffraction efficiency depends only of the product of
these two parameters but, obviously, they are not equivalent
\cite{keller99,champenois01}. Ideally, Bragg diffraction can
provide the $50$\% beam splitters and  $100$\% reflective mirrors
needed to build a perfect Mach-Zehnder interferometer, with a
$100$\% transmission. On the contrary, material gratings have a
low diffraction efficiency in the non-zero orders and the
transmission of this type of interferometer cannot exceed a few
percent \cite{ekstrom93}. Moreover, in the case of the $^2S_{1/2}
-^2P_{3/2}$ transition of lithium, the hyperfine structure of the
excited state is very small. Then, provided that the laser
detuning is large with respect to this structure and that the
laser beam is linearly polarized, the potential seen by ground
state atoms is independent of $M_F$ but still depends of the $F$
value because of the difference in detuning. In this case, the
diffraction amplitude is the same for all the sublevels of one
hyperfine level. For the experiments described below, the laser
frequency is detuned on the blue side of the $^2S_{1/2}-^2P_{3/2}$
transition, the detuning being about $2.1$ and $2.9$ GHz for the
$F=1$ and $F=2$ hyperfine states, respectively. With such a
detuning, real excitation of an atom, while crossing a standing
wave, has a low probability, of the order of a few percent, so
that atomic diffraction is almost perfectly coherent.

A schematic drawing of our interferometer appears in figure 1. An
atomic beam of lithium, strongly collimated by a two-slit system,
crosses three laser standing waves, which play the role of beam
splitters (first and third standing waves) and of mirrors (second
standing wave). Diffraction of an atomic wave of vector ${\mathbf
k}$ by grating $j$ of wavevector $ {\mathbf k}_{gj}$ produces a
new wave of wavevector $ {\mathbf k} + p {\mathbf k}_{gj}$, where
$p$ is the diffraction order. The two waves, exiting from the
interferometer by the exit labeled $1$ in figure 1, have the
wavevectors $({\mathbf k}+{\mathbf k}_{g1} -{\mathbf k}_{g2})$
(upper path) and $({\mathbf k} + {\mathbf k}_{g2} - {\mathbf
k}_{g3})$ (lower path). If these two wavevectors were not equal,
interference fringes would appear on the detector surface and
integration over the detector area would wash out the expected
interference signal. Therefore, we must cancel the quantity
$\Delta{\mathbf k} = {\mathbf k}_{g1} + {\mathbf k}_{g3} - 2
{\mathbf k}_{g2}$ to optimize the fringe contrast. For a perfect
interferometer \cite{champenois99}, the two beams labeled $1$ and
$2$ carry complementary interference signals proportional to the
quantities :

\begin{equation}
I_{1/2} = \left[1 \pm \cos \left(2\pi\left(x_{M1} + x_{M3} -2
x_{M2}\right)/a\right)\right]
\end{equation}

\noindent where $x_{Mi}$ is the $x$ coordinate of mirror $M_i$
producing the laser stationary wave and $a$ is the corresponding
grating period, $a =\lambda_r/2 = 335$ nm. Therefore, interference
fringes can be observed by displacing anyone of the three mirrors
in the $x$-direction.

\begin{figure}
\includegraphics[width=8 cm]{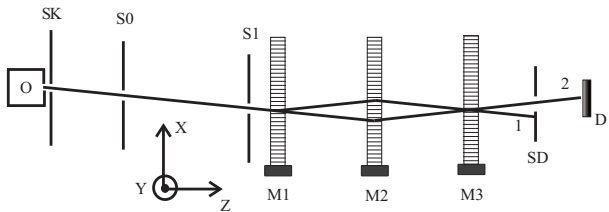}
\caption{\label{drawing} Schematic drawing of a top view of our
interferometer. The $x,y,z$ axis are represented and, for each
element, we give its distance $z$ to the nozzle. O : lithium oven;
Sk : $1$ mm diameter skimmer at $z_s= 15$ mm ; $S_0$ : collimating
slit of width $w_0 = 20$ $\mu$m at $z_{S0} = 480$ mm; $S_1$ :
collimating slit of width $w_1 = 12$ $\mu$m at $z_{S1} = 1260$ mm;
$M_i$, $i=1-3$ : mirrors for the laser standing waves at $z_{M1}
=1410$ mm, $z_{M2} = 2015$ mm and $z_{M3} = 2620 $ mm; $1$ and $2$
: complementary exit beams; $S_D$ : detector slit with tunable
width and x-position at $z_{SD} =3020$ mm ; $D$ : $760$ $\mu$m
wide rhenium ribbon of the Langmuir-Taylor hot wire detector at
$z_D= 3370$ mm. }
\end{figure}

In our experiment, a supersonic beam of lithium seeded in argon
(typical pressure $300$ mbar) is emitted by an oven : the
temperature of the oven body fixes the lithium pressure to $1.6$
mbar at $1023$ K (or $0.65$ mbar at $973$ K for some experiments),
while the front part holding the $200$ $\mu$m nozzle is overheated
($+50$ K) to prevent clogging. To insure the best stability, these
temperatures are stabilized within $\pm1$ K. The beam goes through
a skimmer and enters a second vacuum chamber, where it reaches a
collision-free zone. A third vacuum chamber holds the two-slit
system used to strongly collimate the beam. In the fourth vacuum
chamber, the atomic beam crosses three laser standing waves, each
one being produced by reflecting a laser beam on a mirror $M_i$,
$i=1-3$, the distance between consecutive standing waves being
$605$ mm. One of the emerging atomic beams is then selected by a
detector slit $S_D$ whose width $w_D$ and $x$-position can be
finely tuned under vacuum. Finally, this selected beam enters an
UHV chamber through a $3$ mm diameter hole. This hole and the
skimmer are the only collimating elements in the vertical
$y$-direction. The atoms are detected by a Langmuir-Taylor
hot-wire detector using a rhenium ribbon and a channeltron. The
UHV chamber is pumped by a $100$ l/s turbomolecular pump (base
pressure $10^{-8}$ mbar). The oven chamber is pumped by an
unbaffled $8000$ l/s oil diffusion pump, while all the other
chambers are pumped by oil diffusion pumps, with water cooled
baffles, providing a base pressure near $10^{-7}$ mbar. Our
Langmuir-Taylor hot-wire detector \cite{delhuille02} has a
detection probability for lithium atom in the $10-70$ \% range
depending on rhenium surface oxidation and temperature and a
background count rate of the order of a few thousand counts/s.

The alignment of the collimating elements is simplified thanks to
a laser alignment done before operation. The three mirrors $M_i$
must be oriented within about $20$ microradians, in two directions
and the final adjustments are made under vacuum by piezo mounts.
The properties of a standing wave depend linearly on the
orientation angles of the mirror used to reflect the laser beam,
but are considerably less sensitive to the direction of this beam,
which must be perpendicular to the mirror within $1$ milliradians
only. We use $13$ mm diameter laser beams, produced by splitting
the beam of an argon ion pumped cw single frequency dye laser,
after expansion by a $5\times$ telescope.

\begin{figure}[htb]
\includegraphics[width=8 cm]{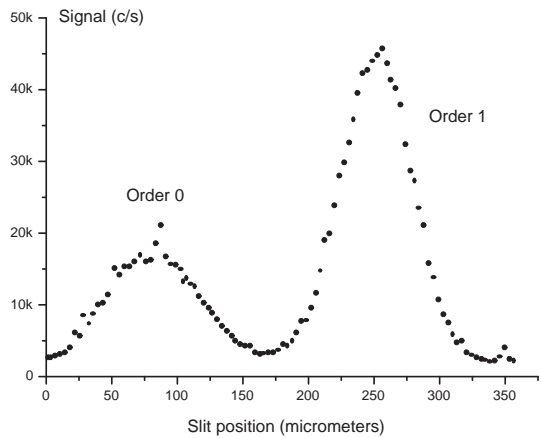}
\caption{\label{diffraction} Laser diffraction of the lithium beam
: number of atoms detected per second as a function of the
x-position (in $\mu m$) of the detector slit $S_D$ (counting
period : $1$ s). Two peaks are observed, corresponding to the
diffraction orders $0$ and $+1$. In this experiment, only one
laser standing wave, associated to mirror $M_2$, is present. We
have verified the absence of a peak corresponding to the order
$-1$, in agreement with the theoretical prediction for Bragg
regime. A few noise bursts of the Langmuir-Taylor hot wire
detector are visible.}
\end{figure}

Figure 2 shows the profile of the lithium beam diffracted by the
laser standing wave corresponding to mirror $M_2$. This profile is
recorded by moving the detector slit $S_D$ (slitwidth $w_D= 50$
$\mu$m). After fine tuning of the angle $\theta_y$ corresponding
to the rotation of this mirror around the y-axis, we observe two
well resolved peaks, corresponding to the diffraction orders zero
and one. The order zero peak contains the undiffracted part of the
$^7$Li $F=1 \mbox{ and } 2$ levels and also the $^6$Li content of
the beam for which the laser has little effect. Bragg diffraction
is a Rabi-type oscillation between the two diffraction orders and
the amplitude and the phase of this oscillation depend of the atom
incidence angle and velocity, so that the observed diffraction
efficiency results from an average over the initial conditions.
>From the geometry of the experiment, we have verified that the
distance between the two peaks is in excellent agreement with the
calculated diffraction angle. The observation of diffraction by
each of the three laser standing waves serves to optimize the
$\theta_y$ angle of each of the three mirrors to fulfill Bragg
condition.

We can then search for interference signals, by running
simultaneously the three standing waves with incident laser powers
equal to $40$, $80$ and $40$ mW respectively, corresponding to an
ideal Mach-Zehnder design. Using an autocollimator, we set the
orientation angles $\theta_z$ of the three mirrors so that the
vector normal to each mirror is horizontal within $50$ $\mu$rad.
As explained above, we must cancel the quantity $ \Delta{\mathbf
k} = {\mathbf k}_{g1} + {\mathbf k}_{g3} - 2 {\mathbf k}_{g2}$ and
this is done by acting on one of the three mirrors, in order to
optimize the fringe contrast.

In several previous apparatuses
\cite{keith91,schmiedmayer97,giltner95b}, the vibrational noise on
the grating $x$-positions induced a large phase noise in the
interferometer, so that it was necessary to measure and reduce
this vibrational noise before any observation. A three-grating
Mach-Zehnder optical interferometer linked to the gratings can be
used for this purpose \cite{keith91,giltner95b}, as first done for
a neutron interferometer by M. Gruber et al. \cite{gruber89}. We
have also built a similar optical interferometer. Its output
signal is given by equation (1), $a$ being now the optical
gratings period ($a = 5$ $\mu$m in our experiment). In our
interferometer, the detected part $(x_{M1} + x_{M3} -2 x_{M2})$ of
the vibration-induced motion of the three mirrors has a rms value
equal to $3$ nm in a $50$ kHz bandwidth. This very small
vibrational noise is due to our construction (the interferometer
mirrors are on a very rigid bench inside the vacuum chambers,
which are placed on a massive support located in the basement). As
the resulting phase noise, $6\times10^{-2}$ rad, induces a
negligible contrast loss, we have not tried to reduce this noise
by a feedback loop.

\begin{figure}
\includegraphics[width=8 cm]{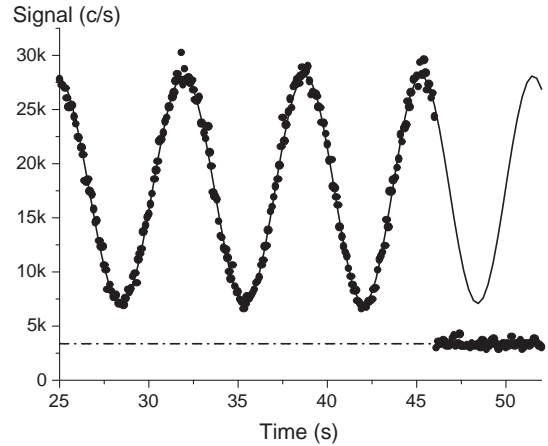}
\caption{\label{interference} Interference fringes : number of
atoms detected per second as a function of time (counting period :
$0.1$ s). The position $x_{M3}$ of mirror $M_3$ is swept as a
function of time. We have verified that the fringe period
corresponds to a displacement $\Delta x_{M3} = \lambda_r/2= 335$
nm. The background signal of the detector, measured by flagging
the lithium beam $50$ s later, is shown on the right part of the
figure and its average value $3370$ counts/s is represented by the
dot-dashed line. The data points are fitted by a sine curve, whose
argument is the sum of linear and a quadratic functions of time
(this last term being needed to represent the piezo hysteresis).
>From this fit, we extract the value of the fringe contrast equal
to $74$\% with an error of the order of $1$\%.}
\end{figure}

The detector slit $S_D$, with a width $w_D= 30$ $\mu$m, has been
put at exit $1$ or at exit $2$ (see figure 1) with similar
results. A slightly better fringe contrast is obtained at exit $2$
than at exit $1$, probably because of different contributions of
stray atomic beams in the two cases. The main stray beams, which
are not represented in figure 1, correspond to the neglected
diffraction orders : they should vanish exactly if the laser power
densities and the interaction times were perfectly tuned and if
there was no angular and velocity dispersion of the incident
atomic beam. Figure 3 presents an example of experimental signal
collected at exit $2$ with a counting period equal to $0.1$ s. If
we substract the background which has an average value of $3370$
counts/second, we estimate the fringe contrast by :

$${\mathcal{C}} = (I_{max} -I_{min})/(I_{max} +I_{min})\approx
0.74 $$

\noindent A simulation of our interferometer (as in our paper
\cite{champenois99}, but using the Bragg diffraction amplitudes
corresponding to an ideal interferometer) predicts a contrast near
$90$\%, limited by the overlap of the exit beams $1$ and $2$ (see
figure 1). The difference between this simulation and our
experiment is parly due to stray beams, partly due to some phase
dispersion in the interferometer. Assuming that the dominant
effect is due to phase dispersion, with a Gaussian distribution
and a rms value $\sigma$, the contrast is reduced by the factor
$\exp(-\sigma^2/2)$. We thus deduce $\sigma\approx0.63$ rad : in
the language of usual optics, the rms value of the wavefront
defects is equal to $\lambda/10$, where $\lambda$ is the atom
wavelength close to $54$ picometer!

Finally, we have measured the phase sensitivity of our
experimental signal near $17$ mrad$/\sqrt{{\mbox Hz}}$, not far
from the shot-noise limit $\approx 10$ mrad$/\sqrt{{\mbox Hz}}$
deduced from the signal and background count rates.

As a conclusion, we have built and operated a Mach-Zehnder Bragg
atom interferometer with lithium and obtained first interference
signals with an excellent signal to noise ratio and a high fringe
contrast, equal to $74$ \%. Our simulations indicate that the
present fringe contrast can be improved and we expect to do so in
a near future. The contrast we have observed is comparable to the
contrast observed with most cold atom interferometers (for
instance \cite{kasevich91,peters99}). However, a contrast of
nearly $100$\% has been achieved in a Mach-Zehnder Bragg
interferometer using a Bose-Einstein condensate as the atomic
source \cite{torii00}. It is also interesting to compare our
results with neutron interferometers. The technique to build such
interferometers is now mature and recently built neutron
interferometers \cite{gilliam99,kroupa00} exhibit a fringe
contrast near $90$\%, while a $68$\% contrast was already observed
in 1978 \cite{bauspiess78}. These very nice results suggest that
an extremely high contrast is feasible. Unfortunately, the
interactions of neutrons and atoms with matter and fields are
extremely different so that the know-how established with neutrons
is not easily transferred to atom interferometers.

We expect to optimize our setup, in particular to increase the
beam intensity by various means, including transverse laser
cooling. It will also be possible to work separately with both
lithium isotopes, a very interesting possibility for some
applications. In our interferometer, the two atomic paths are
separated by $100$ $\mu$m near the second standing wave and this
distance is substantially larger than in previous atom
interferometers and even larger separations have been achieved by
the group of Toennies \cite{toennies01}. With this new generation
of interferometers, very sensitive measurements of weak
perturbations are possible : with an interaction time $\tau$ of
the order of $100$ $\mu$s and a minimum detectable phaseshift of
the order of $0.1$ mrad (which seems within reach, after some
optimization, with an integration time of the order of a few
hours), a perturbation as small as $6\times10^{-16}$ eV can be
detected. This extreme sensitivity will be used to measure atomic
polarizability, index of refraction of gases for atomic waves,
following the previous works of Pritchard's group
\cite{schmiedmayer97,ekstrom95,schmiedmayer95}.

\section{acknowledgement}

We are very much indebted to the technical staff of our
laboratory, M. Gianesin, D. Castex, P. Paquier, T. Ravel, A.
Pellicer, L. Polizzi, W. Volondat, for their help in building the
interferometer and to several students, A. Miffre, Th. Lahaye, E.
Lavallette, R. Saers, B. Aymes, J. de Lataillade, for their
contribution to the experiments. It is a pleasure to thank various
colleagues, A. Bordenave-Montesquieu, J. F. Fels, J. P. Gauyacq,
Siu Au Lee, H. J. Loesch, J. P. Toennies and J. P. Ziesel for
their help and advice, and the " P\^ole Optique et Vision " (Saint
Etienne) for laser machining of our collimating slits. R\'egion
Midi Pyr\'en\'ees is gratefully acknowledged for financial
support. We also thank CNRS/SPM for financial support and for a
temporary position given to one of us (L.J.).

%%%%%%%%%%%%%%%%%%%%%%%%%%%%%%%%%%%%%%%%%%%%%%%%%%%%%%%%%%%%%%%%%%%%

\end{document}